\documentclass[12pt,preprint]{aastex}

\newcommand{\CHAX}{CH~A$^{2}\Delta$--X$^{2}\Pi$}
\newcommand{\CHBX}{CH~B$^{2}\Sigma^{-}$--X$^{2}\Pi$}
\newcommand{\BX}{B$^{2}\Sigma$--X$^{2}\Sigma$}
\newcommand{\CNBX}{CN~B$^{2}\Sigma^{+}$--X$^{2}\Sigma^{+}$}

\newcommand{\FeI}{\ion{Fe}{1}}
\newcommand{\CI}{\ion{C}{1}}
\newcommand{\OI}{\ion{O}{1}}

\usepackage{natbib}

\shortauthors{Tritschler, A. \& Uitenbroek, H.}
\shorttitle{}

\newcommand{\aliemail}{ali@bbso.njit.edu}
\newcommand{\hanemail}{huitenbroek@nso.edu}

\begin{document}

%
%

\title{The contrast of magnetic elements in synthetic CH- and CN-band
images of solar magnetoconvection}
\author{H.~Uitenbroek}
\affil{National Solar Observatory/Sacramento Peak\footnote{Operated by the %
       Association of Universities for Research in Astronomy, Inc. (AURA), %
       for the National Science Foundation},
       P.O.~Box 62, Sunspot, NM-88349, U.S.A.}
\email{\hanemail}
\author{A.~Tritschler}
\affil{Big bear Solar Observatory,
       New Jersey Institute of Technology,
       40386 North Shore Lane,
       Big bear City, CA-92314, U.S.A.}
\email{\aliemail}


%
%

\begin{abstract}
We present a comparative study of the intensity contrast in synthetic
CH-band and violet CN-band filtergrams computed from a high-resolution
simulation of solar magnetoconvection.
The underlying simulation has an average vertical magnetic field of
250\,G with kG fields concentrated in its intergranular lanes,
and is representative of a plage region.
To simulate filtergrams typically obtained in CH- and CN-band
observations we computed spatially resolved spectra in both bands
and integrated these spectra over 1\,nm FWHM filter functions 
centred at 430.5\,nm and 388.3\,nm, respectively.
We find that the average contrast of magnetic bright points in the
simulated filtergrams is lower in the CN-band by a factor of 0.96.
This result strongly contradicts earlier semi-empirical modeling
and recent observations, which both etimated that the bright-point
contrast in the CN-band is \emph{higher} by a factor of 1.4.
We argue that the near equality of the bright-point contrast
in the two bands in the present simulation is a natural
consequence of the mechanism that causes magnetic flux elements
to be particularly bright in the CN and CH filtergrams,
namely the partial evacuation of these elements and the
concomitant weakening of molecular spectral lines in the
filter passbands.
We find that the RMS intensity contrast in the whole field-of-view of
the filtergrams is 20.5\% in the G band and 22.0\% in the CN band and
conclude that this slight difference in contrast is caused by the
shorter wavelength of the latter.
Both the bright-point and RMS intensity contrast in the CN band are
sensitive to the precise choice of the central wavelength of the filter.
\end{abstract}


%
%

\section{Introduction}\label{sec:intro}
Morphological and dynamical studies of small-scale magnetic 
flux concentrations on the solar surface are challenged 
by short evolutionary time scales, and spatial scales that are
close to the diffraction limit of most solar telescopes,
even those with large apertures.
As a result magnetograms often lack the necessary 
spatial and/or temporal resolution to allow adequate identification
and tracing of these magnetic features.
In this context broad-band imaging in molecular bands towards the blue
end of the solar optical spectrum greatly contributed to our current
understanding of the smallest manifestations of solar magnetic flux.
High-spatial resolution filtergram observations in the notorious G band
around 430.5\,nm
	\citep{Muller+Hulot+Roudier1989,Muller+Roudier1992,%
Muller+Roudier+Vigneau+Auffret1994,Berger_etal1995,%
VanBallegooijen_etal1998,Berger+Title2001}
show high contrasted (typically 30\,\%) subarcsecond sized 
brightenings embedded in intergranular lanes
        \citep{Berger_etal1995}.
        \citet{Berger+Title2001} 
found that these G-band bright points are cospatial and comorphous
with magnetic flux concentrations to within 0.24\,arcsec. 

The G-band region is mostly populated by electronic transitions
in the \CHAX\ molecular band.
A similar band results from \BX\ transitions of the CN molecule
at 388.3\,nm.
Several authors have suggested that because of its shorter wavelength
and a correspondingly higher Planck sensitivity
the contrast of magnetic elements in CN-band filtergrams
could be more pronounced, making the latter an even more attractive
magnetic flux proxy.
Indeed, the relative brightness behaviour in the two molecular bands
in semi-empirical fluxtube models
      \citep{Rutten+Kiselman+Rouppe+Plez2001}
and Kurucz radiative equilibrium models of different effective
temperature
      \citep{Berdyugina+Solanki+Frutiger2003}
strongly points in this direction.
Observational evidence in support of such promising semi-empirical
estimates was found by 
      \citet{Zakharov+Gandorfer+Solanki+Loefdahl2005}
based on reconstructed simultaneous images in the G band and the CN band
obtained with adaptive optics support at the 1-m Swedish Solar Telescope
(SST) on La Palma.
These authors concluded that their observed bright-point contrast was
typically 1.4 times higher in the CN band than in the G band.

In order to verify/illuminate the aforementioned
suggestion in a more realistic solar model we compare 
the contrast of solar magnetic elements in
synthetic CH- and CN-band filtergrams computed from a snapshot of solar
magnetoconvection to determine which would be more suitable for
observations at high spatial resolution.
Similar modeling was performed by
     \citet{Schuessler_etal2003}
to investigate the mechanism by which magnetic elements appear
bright in G-band filtergrams, and by
     \citet{Carlsson+Stein+Nordlund+Scharmer2004}
to study the center-to-limb behaviour of G-band intensity in small-scale
magnetic elements.
Much earlier, the CN- and CH-bands have been modelled extensively by 
     \citet{Mount+Linsky+Shine1973,Mount+Linsky1974a,Mount+Linsky1974b,%
Mount+Linsky1975a,Mount+Linsky1975b} to investigate the
thermal structure of the photosphere in the context of one-dimensional
hydrostatic modeling.

Because broad-band filters integrate in wavelength
and average over line and continuum intensities,
images obtained with them would seem, at first sight,
not very well-suited for a detailed comparison between
observations and numerical simulations.
Yet, because of the high spatial resolution that can be
achieved in broad-band filtergrams, and precisely because
the filter signal only weakly depends on the properties of
individual spectral lines, such images make ideal targets
for a comparison with numerical simulations.
Properties like the average intensity contrast through the filter,
the average contrast of bright points, and the relative behaviour
of these contrasts at different wavelengths are a corollary
of the present computations and can be compared
in a statistical sense with observations to assess the realism
of the simulations.

We summarise the spectral modeling in Section \ref{sec:synthesis},
introduce intensity response functions as a way to estimate the
formation height of filter intensities in Section \ref{sec:response}, 
and present results for the bright-point contrasts in
Section \ref{sec:contrast}. 
The results are discussed and concluded in Sections 
\ref{sec:discussion} and Sections \ref{sec:conclusion}, respectively.

%
%

\section{Spectral synthesis}\label{sec:synthesis}
To investigate the relative behaviour of bright-point contrast
in the CH-line dominated G band and the CN band at 388.3\,nm
we synthesised the emergent intensities at both wavelengths
through a snapshot from a high-resolution magnetoconvection
simulation containing strong magnetic fields
      \citep{Stein+Nordlund1998}.
Magnetoconvection in this type of simulation is realized after
a uniform vertical magnetic seed field with a flux density of 250\,G
is superposed on a snapshot of a three-dimensional hydrodynamic
simulation and is allowed to develop. 
As a result the magnetic fields are advected to the mesogranular boundaries
and concentrated in downflow regions showing field strengths
up to 2.5\,kG at the $<\tau_{500}> = 1$ level.
The simulation covers a small 6$\times$6\,Mm region of the solar
photosphere with a 23.7\,km horizontal grid size,
and spans a height range from the temperature minimum at around 0.5\,Mm to
2.5\,Mm below the visible surface, where $z=0$ corresponds to
$<\tau_{500}>=1$. 
Given its average flux density the employed simulation is
representative of plage, rather than quiet Sun.
To account for the interaction between convection and radiation the
simulations incorporate non-gray three-dimensional radiation transfer
in Local Thermodynamic Equilibrium (LTE) by including a radiative
heating term in the energy balance and LTE ionization and excitation
in the equation of state.
For the radiative transfer calculations presented here the vertical
stratification of the simulation snapshot was re-interpolated to a
constant grid spacing of 13.9\,km with a depth extending to 300\,km
below the surface from the original resolution of 15\,km in the upper
layers to 35\,km in the deep layers.
The same snapshot has been used by
     \citet{Carlsson+Stein+Nordlund+Scharmer2004}
to study the center-to-limb behaviour of faculae in the G band.

\subsection{Molecular number densities\label{sec:densities}}
The coupled equations for the concentrations of the molecules
H$_2$, CH, CN, CO and N$_2$, and their constituent atoms were solved
under the assumption of instantaneous chemical equilibrium
      \citep[e.g.,][p.\ 46]{AAQ_4}.
To solve for such a limited set of molecules
is justified because only a small fraction of the atoms
C, N and O is locked up in molecules other than the five we considered.
In a test calculation with a two-dimensional vertical slice
through the data cube we found that the CN and CH concentrations
deviated only by up to 0.15\,\% and 0.2\,\%, respectively,
from those calculated with a larger set of 12 of the most abundant
molecules, including in addition H$_2^+$, NH, NO, OH, C$_2$, and H$_2$O.
Dynamic effects are not important for the disk centre intensities
we calculate
     \citep{AsensioRamos+TrujilloBueno+Carlsson+Cernicharo2003,%
WedemeyerBoehm+Kamp+Bruls+Freytag2005}

We used a carbon abundance of $\log \epsilon_{C} = 8.39$
as advocated by
     \citet{Asplund+Grevesse+Sauval+AllendePrieto+Blomme2005}
on the basis of \CI, CH, and C$_2$ lines modelled in three-dimensional
hydrodynamic models, and an oxygen abundance of $\log \epsilon_{O} = 8.66$
as determined from three-dimensional modeling of \OI, and OH lines by
     \citet{Asplund+Grevesse+Sauval+AllendePrieto+Kiselman2004}.
This carbon abundance is in good agreement with the value
of $\log \epsilon_{C} = 8.35$ on the basis of analysis
of the same CN violet band we consider here
     \citep{Mount+Linsky1975b}.
We assume the standard nitrogen abundance of $\log \epsilon_{N} = 8.00$ of
     \citet{Grevesse+Anders1991}.
Dissociation energies of $D_0 = 3.465$\,eV for CH and and 7.76\,eV for CN,
and polynomial fits for equilibrium constants and partition functions
were taken from
     \citet{Sauval+Tatum1984}.

%
%
\begin{figure}[hbtp]
  \epsscale{0.7}
  \plotone{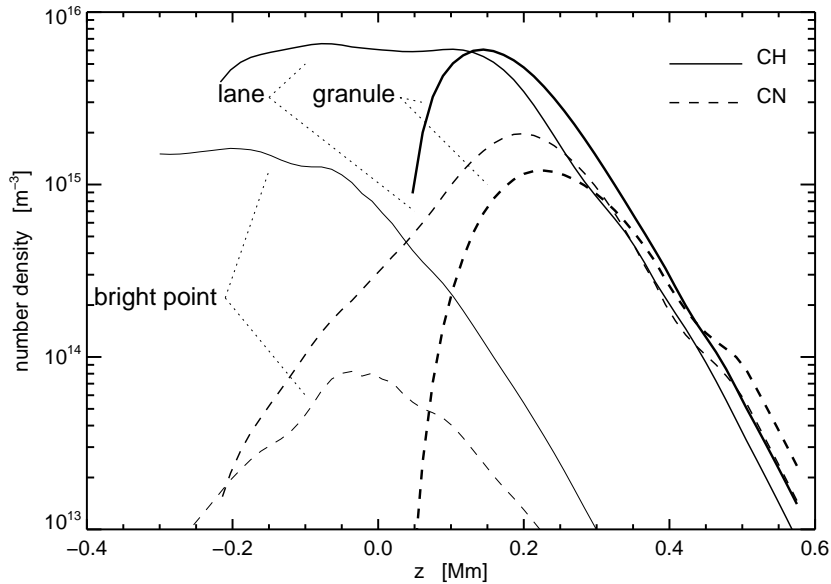}
  \caption{Number density of CH and CN molecules in 
           a granule, an intergranular lane, and a magnetic flux element.
           \label{fig:concentrations}}
  \epsscale{1.0}
\end{figure}
For comparison the number densities of CH and CN in the snapshot
are shown in Figure \ref{fig:concentrations} as a function of height $z$
in three characteristic structures: a granule, a weakly magnetic
intergranular lane, and a magnetic element with strong field.
Because hydrogen is more abundant than nitrogen the density
of CH is generally higher than that of CN.
While the ratio of number densities is about a factor of 2--5
in the middle photosphere, the difference is much larger in
deeper layers, and is slightly reversed in the topmost layers.
The strong decline in CN number density in deeper layers
is the result of the temperature sensitivity of
the molecular association-dissociation equilibrium,
which is proportional to $\exp(D_0/kT)$ with the dissociation
energy $D_0$ of CN twice that of CH.

In the magnetic concentration internal gas pressure plus
magnetic pressure balances external pressure.
At a given geometric height, therefore, the internal gas pressure and
the density are lower in the flux element compared to its surroundings:
it is partially evacuated.
As a result the molecular density distributions in the flux concentration
appear to be shifted downward by about 250\,km with
respect to those in the weakly magnetic intergranular lane.
Moreover, because the magnetic field at a given height 
in the magnetic element in part supports the gas column above
that height the gas pressure is lower than it is at the same
temperature in the surroundings.
Therefore, partial pressures of the molecular constituents
are lower and, through the chemical equilibrium equation,
this leads to a lowering of the molecular concentration
curves in addition to the apparent shift
      \citep[see also][]{Uitenbroek2003}.

\subsection{Spectra\label{sec:spectra}}
Spectral synthesis of the molecular bands was accomplished in
three-dimensional geometry with the transfer code RHSC3D, 
and in two-dimensional vertical slices of the
three-dimensional cube with the code RHSC2D.
These are described in detail in
     \citet{uitenbroek1998, uitenbroek2000a, uitenbroek2000b}. 
For a given source function the radiation transfer 
equation was formally solved using the short characteristics method 
     \citep{kunasz+auer1988}.
All calculations were performed assuming LTE source functions and
opacities.

The emergent spectra in the vertical direction were calculated for
two wavelength intervals of 3 nm width centered on 388.3 nm at
the CN band head, and at 430.5 nm in the G band, respectively
(all wavelengths in air).
In each interval 600 wavelength points were used.
This fairly sparse sampling of the wavelength bands is dense
enough for the calculation of the wavelength integrated filter signals
we wish to compare.
We verified the accuracy of the derived filter signals by comparing
with a calculation that uses 3000 wavelength points in each interval
in a two-dimensional vertical slice through the snapshot cube,
and found that the RMS difference between the filter signal derived
from the dense and the coarse wavelength sampling was only 2\%.

Line opacities of atomic species and of the CN and CH molecules in the
two wavelength intervals were compiled from
      \citet{Kurucz_CD13,Kurucz_CD18}.
Voigt profiles were used for both molecular and atomic lines
and these were calculated consistently with temperature and
Doppler shift at each depth.
No micro- or macro-turbulence, nor extra wing damping was
used as the Doppler shifts resulting from the convective
motions in the simulation provide realistic line broadening.
To save on unnecessary Voigt function generation we eliminated
weak atomic lines from the line lists and kept 207 relevant atomic
lines in the CN band and 356 lines in the G band interval.
The CN band wavelength interval includes 327 lines of the
\CNBX\ system ($v = 0 - 0$, where $v$ is the vibrational quantum
number) from the blue up to the band head proper at 388.339\,nm.
This interval also contains many weak lines ($gf \leq -5$) of the
\CHAX\ system (231 lines with $v = 0 - 1$ and $v = 1 - 2$),
and 62 stronger lines of the \CHBX\ system ($v = 0 - 0$),
in particular towards the red beyond $\lambda = 389$\,nm.
A dominant feature in the red part of the CN band wavelength
interval is the hydrogen Balmer line H$_8$ between levels
$n = 8$ and 2 at $\lambda = 388.905$\,nm.
This line is not very deep but has very broad damping wings.
The wavelength interval for the G band includes 424 lines
of the \CHAX\ system with $v = 0 - 0, 1 - 1$, and $2 - 2$.
%
%
\begin{figure}[hbtp]
  \epsscale{0.9}
  \plotone{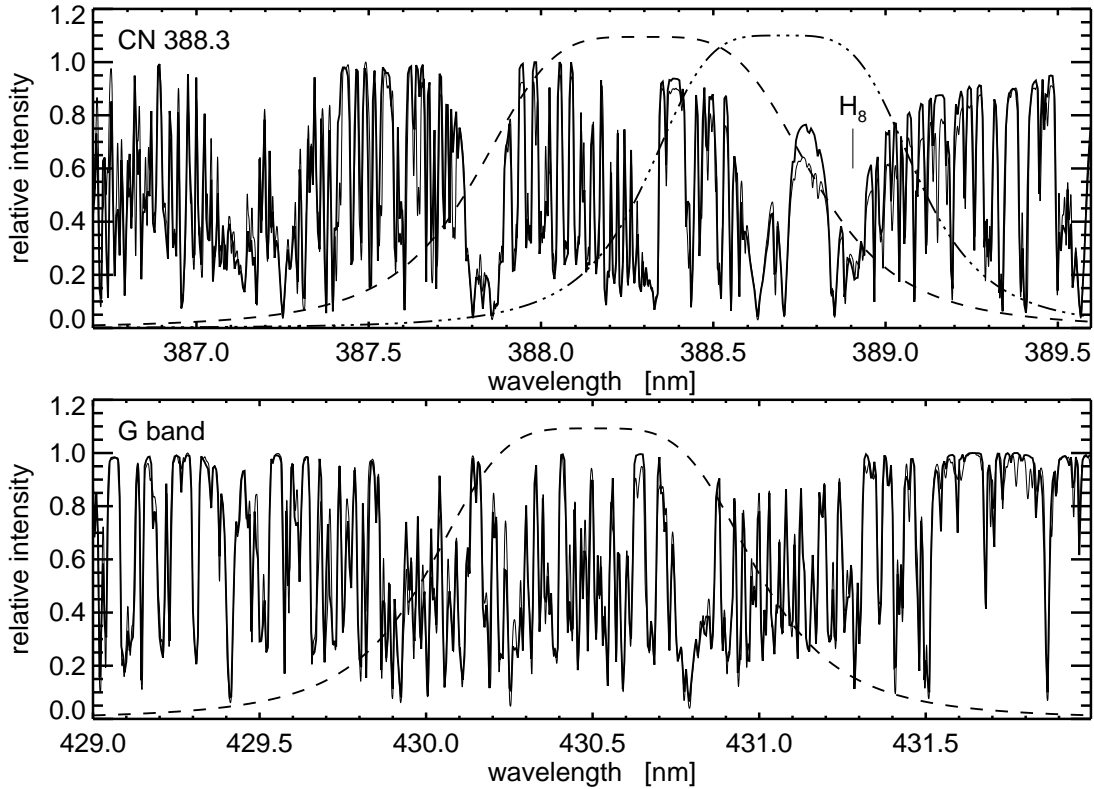}
  \caption{Spatially averaged emergent spectra in the CN band
           (top panel) and G-band intervals (thick curves).
           Thin curves show the disk-center atlas spectrum for
           comparison.
           The filter functions we employed are drawn with the 
           dashed curves in both panels, while the curve of the filter
           corresponding to the one employed by Zakharov et al.\ is
           indicated by the dot-dashed curve.
           The position of the hydrogen Balmer H$_8$ line is
           marked in the top panel near $\lambda = 388.905$\,nm.}
  \label{fig:spectrum}
  \epsscale{1.0}
\end{figure}
%

\begin{table}[h]
  \caption{Parameters of the CH and CN band filters.
           \label{tab:filters}}
  \vspace*{2ex}
  \begin{tabular}{lcc}
     \hline\hline
     Filter        &   $\lambda_0$ [nm]   & $\lambda_{\mathrm{FWHM}}$ \\ \hline
     G-band        &     430.5       &  1.0 \\
     CN            &     388.3       &  1.0 \\
     CN (Zakharov) &     388.7       &  0.8 \\
     CN (SST) &     387.5       &  1.0 \\ \hline
  \end{tabular}
\end{table}
The emergent spectra in the two intervals, averaged over the surface
of the three-dimensional snapshot and normalised to the continuum,
are shown in Figure \ref{fig:spectrum} and are compared to a
spatially averaged disk-centre atlas
      \citep{Brault+Neckel1987,Neckel1999}.
The calculated spectra are in excellent agreement with the atlas,
and confirm the realism of the simulations and the spectral synthesis.
Also drawn in Figure \ref{fig:spectrum} are the CN and CH filter
functions we used (dashed lines).
The employed filter curves are generalised Lorentzians of the form:
\begin{equation}
  F_{\lambda} = \frac{T_{\mathrm{max}}}%
{1 + \left\{\frac{2(\lambda - \lambda_0)}%
{\lambda_{\mathrm{FWHM}}}\right\}^{2n}}
  \label{eq:filterfunction}
\end{equation}
with order $n = 2$, representative of a dual-cavity interference filter.
In eq.\ [\ref{eq:filterfunction}] $\lambda_0$ is the filter's central
wavelength, $\lambda_{\mathrm{FWHM}}$ is its width at half maximum,
and $T_{\mathrm{max}}$ is its transmission at maximum.
We list the parameters of our filter functions with values
typically used in observations, in the first two rows of
Table \ref{tab:filters}.
In addition, we list the parameters for the filter used by
     \citet{Zakharov+Gandorfer+Solanki+Loefdahl2005},
and the filter listed on the support pages of the Swedish Solar Telescope
(SST) on La Palma.

The broad-band filter used by
     \citet{Zakharov+Gandorfer+Solanki+Loefdahl2005}
to investigate the brightness contrast in the CN band in comparison with
the G-band is centered at $\lambda_0 = 388.7$\,nm, redward of the CN
band head at 388.339\,nm.
Curiously, it receives only a very small contribution from CN lines
because of this.
The filter mostly integrates over three \FeI\ lines at $\lambda$
388.629\,nm, 388.706\,nm and 388.852\,nm, the Balmer H$_8$ line,
and the CH lines around 389\,nm.
For comparison, the estimated transmission function for this filter
is drawn with the dash-dotted curve in the top panel of
Figure \ref{fig:spectrum}.
The G-band filter used by these authors has the same parameters as
the one used in the theoretical calculations presented here.

\subsection{Synthetic filtergrams\label{sec:images}}
\begin{figure}[tbhp]
  \plottwo{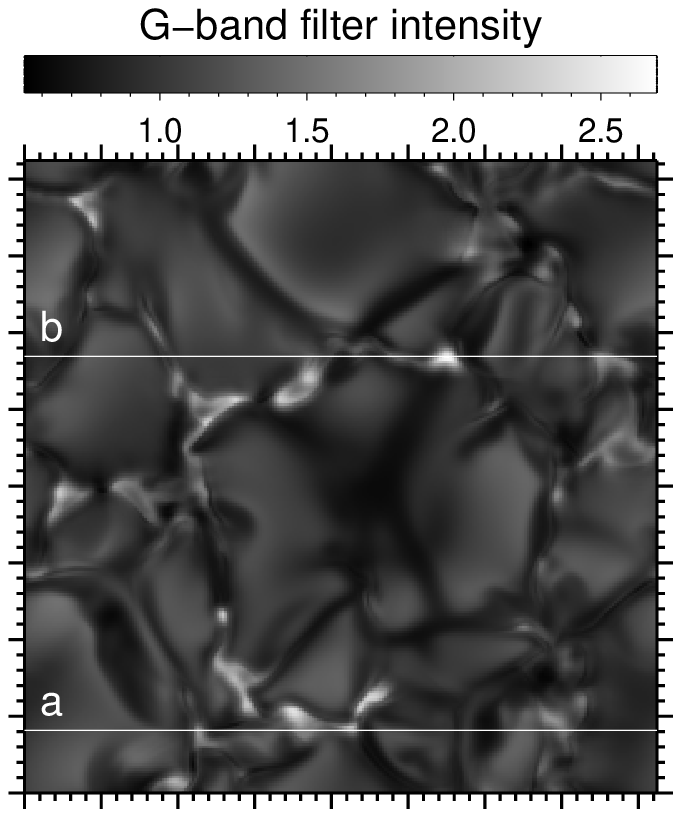}{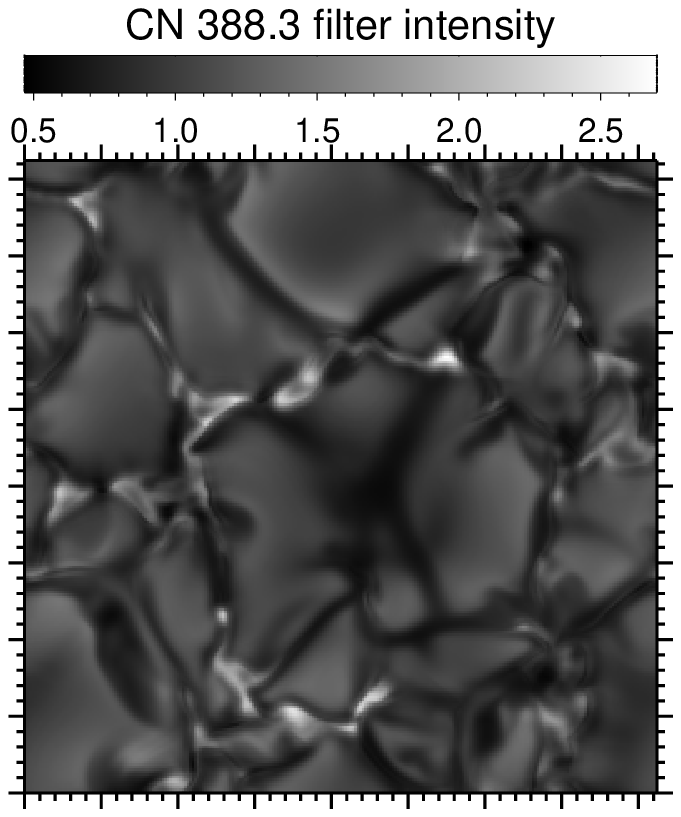}
  \caption{Synthetic filtergrams in the G band (left panel) and violet CN
           band constructed from the calculated disk-centre spectra.
           The intensity in each filtergram is normalised with respect
           to the average quiet-Sun value for that band.
           Major tick marks correspond to one arcsec intervals.
           Horizontal lines $a$ and $b$ mark cross sections used
           in Figures \ref{fig:response} and \ref{fig:cuts}.
           \label{fig:maps}}
\end{figure}

Based on the calculated disk-centre spectra we synthesise filtergrams 
by taking into account the broad-band filters specified 
in Table \ref{tab:filters} and intergrating over wavelength. 
Figure \ref{fig:maps} presents the result for the G-Band (left panel)
and the CN band (right panel).
The filtergrams look almost identical with each showing very clearly
the bright point and bright elongated structures associated with
strong magnetic field concentrations.
The filtergrams were normalised to the average quiet-Sun intensity
in each passband, defined as the spatial averaged signal for all
pixels outside the bright points (see Sect. \ref{sec:contrast}).
%
%
%

The RMS contrast over the whole field-of-view (FOV) is 22.0\,\% and
20.5,\% for the CN band and the G band, respectively.
The larger contrast in the CN band is the result of its shorter wavelength,
but the difference is much smaller than expected on the basis of
consideration of the temperature sensitivity of the Planck function.
A convenient measure to express the difference in temperature
sensitivity of the Planck fuction $B_{\lambda}(T)$
between the two wavelengths of the molecular bands is the ratio
\begin{equation}
  \frac{B_{388.3}(T)}{B_{388.3}(4500)} / 
    \frac{B_{430.5}(T)}{B_{430.5}(4500)},
  \label{eq:Bratio}
\end{equation}
where $T=4500$ is the average temperature of the photosphere at $z = 0$.
Since characteristic temperature differences between granules and
intergranular lanes are 4000\,K at this height we would expect
a much higher value for the ratio of the granular contrast bewteen
the CN and the CH band (eq. [\ref{eq:Bratio}] gives 1.26 for $T = 6500$,
and 1.45 for $T = 8500$) than the one we find in the present filtergrams.
However, three circumstances reduce the contrast
in the 388.3\,nm filter signal with respect to that in the 430.5\,nm band. 
First, the filter averages over lines and continuum wavelengths
and at the formation height of the lines the temperature fluctuations
are much smaller (e.g., at $z = 150$\,km the temperature differences
are typically only 800\,K).
Secondly, because of the strong temperature sensitivity of the H$^-$
opacity, optical depth unity contours (which provide a rough estimate
of the formation height, see also Sect.\ \ref{sec:response})
approximately follow temperature contours and thus sample much
smaller temperature variations horizontally than they would
at a fixed geometrical height.
Finally, the CN concentration is reduced in the intergranular lanes
(see Figure \ref{fig:concentrations}) with respect to the CH
concentration.
This causes weakening of the CN lines and raises the
filter signal in the lanes, thereby preferentially reducing the
contrast in the CN filtergram compared to values expected from
Planck function considerations.

%
%

\section{Filter Response functions}\label{sec:response}
We use response functions to examine the sensitivity of the filter
integrated signals to temperature at different heights in the solar
atmosphere.
The concept of response functions was first explored by
     \citet{Beckers+Milkey1975}
and further developed by
     \citet{LandiDeglinnocenti+LandiDeglinnocenti1977}
who generalised the formalism to magnetic lines,
and also put forward the {\it line integrated
response function} (LIRF) in the context of broad-band measurements.
We derive the temperature response function in the inhomogeneous
atmosphere by numerically evaluating the changes in the
CH and CN filter integrated intensities that results from
temperature perturbations introduced at different heights in the atmosphere.
Since it is numerically very intensive the computation is performed
only on a two-dimensional vertical cross-section
through the three-dimensional magnetoconvection snapshot,
rather than on the full cube.
Our approach is very similar to the one used by
      \citet{Fossum+Carlsson2005}
to evaluate the temperature sensitivity of the signal observed
through the 160\,nm and 170\,nm filters of the TRACE instrument.

Given a model of the solar atmosphere we can calculate the
emergent intensity $I_{\lambda}$ and fold this spectrum through
filter function $F_{\lambda}$ to obtain the filter integrated
intensity
\begin{equation}
  f = \int_{0}^{\infty} I_{\lambda} F_{\lambda} d \lambda.
  \label{eq:filter}
\end{equation}
Let us define the response function $R^{f,T}_{\lambda}(h)$ of
the filter-integrated emergent intensity to changes in temperature $T$ by:
\begin{equation}
  f \equiv \int_{-\infty}^{z_0} R^{f,T}_{\lambda}(z) T(z) d z,
  \label{eq:response}
\end{equation}
where $z$ is the height in the semi-infinite atmosphere and $z_0$
marks its topmost layer.
Written in this way the filter signal $f$ is a mean representation
of temperature $T$ in the atmosphere weighted by the response function $R$. 
If we now perturb the temperature in different layers in the atmosphere
and recalculate the filter-integrated intensity we obtain a
measure of the sensitivity of the filter signal
to temperature at different heights.
More specifically, if we introduce a temperature perturbation of the form
      \citep{Fossum+Carlsson2005}
\begin{equation}
  \Delta T(z') = t(z') H(z' - z),
  \label{eq:stepfunction}
\end{equation}
where $H$ is a step function that is 0 above height $z$ and 1 below,
the resulting change in the filter-integrated intensity is formally
given by:
\begin{equation}
  \Delta f_z = \int_{-\infty}^{z} R^{f,T}(z') t(z') d z'.
\end{equation}
Subsequently applying this perturbation at each height $z$
in the numerical grid, recalculating $f$, and subtracting the result
from the unperturbed filter signal yields a function $\Delta f_z$ which
can be differentiated numerically with respect to $z$ to recover the
response function:
\begin{equation}
    R^{f, T}(z) = \frac{1}{t(z)} \frac{d}{d z} \left(
      \Delta f_z \right).
\end{equation}
To evaluate the response functions presented here we used perturbation
amplitudes of 1\% of the local temperature, i.e.\ $t(z) = 0.01\quad T(z)$.
Note that we do not adjust the density and ionization equilibria
in the atmosphere when the temperature is perturbed, 
so that the perturbed models are not necessarily physically
consistent.
However, since we only introduce small perturbations, the resulting
error in the estimate of the response function is expected to be small.

\begin{figure}[htbp]
  \epsscale{0.9}
  \plotone{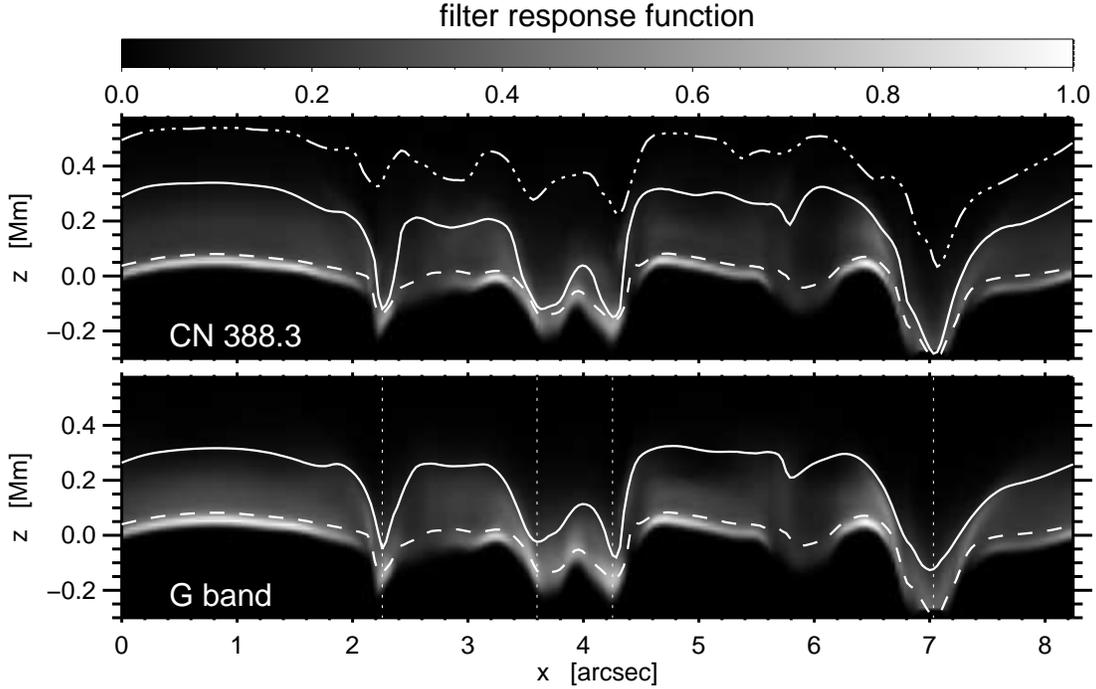}
  \caption{Response functions to temperature for the G-band (bottom)
           and CN-band filter signals in the two-dimensional vertical
           cross section marked $a$ in Figure \ref{fig:maps}.
           Dashed and solid curves mark vertical optical depth unity in
           the continuum and a typical molecular line
           ($\lambda = 387.971$\,nm for CN and $\lambda = 430.390$\,nm
           for CH), respectively.
           Optical depth unity of the CN band head at 388.339\,nm
           is marked with the dash-dotted curve in the top panel.
           The locations of bright points in the cross section
           are indicated by the vertical dotted lines.
	   \label{fig:response}}
  \epsscale{1.0}
\end{figure}

Figure \ref{fig:response} illustrates the behaviour of the
G-band (bottom panel) and CN band head (top panel) filter response
functions $R^{f,T}$ in the inhomogeneous magnetoconvection dominated
atmosphere.
It shows the depth-dependent response function for the two filter
intensities in the vertical slice through the simulation snapshot
marked by $a$ in the G-band panel in Figure \ref{fig:maps}.
This cut intersects four G-band (and CN band) bright points at
$x = 2.2, 3.6, 4.2$, and 7.0 arcsec, the location of which is marked
by the vertical dotted lines in the bottom panel.
The solid and dashed curves mark the location of optical depth
$\tau_l = 1$ in the vertical line-of sight in a representative
CN and CH line core, and optical depth $\tau_c = 1$ in the continuum
in each of the two wavelength intervals, respectively.
The dash-dotted curve in the top panel marks optical depth $\tau_h = 1$
in the CN band head at 388.339\,nm.

The response functions have their maximum for each position along
the slice just below the location of optical depth unity in the continuum
at that location, indicating that the filter intensity is most sensitive
to temperature variations in this layer.
At each $x$ location the response functions show an upward extension
up to just below the $\tau_{l} = 1$ curves.
This is the contribution of the multitude of molecular and atomic
lines to the temperature sensitivity of the filter signals.
We note that both the CN and CH filter response functions are very
similar in shape, vertical location, and extent, with a slightly
larger contribution of line over continuum in the case
of the G band, which is related to the larger number densities of
CH (see Figure \ref{fig:concentrations}).
The highest temperature sensitivity results from the large continuum
contribution over the tops of the granules.
This is where the temperature gradient is steepest and the lines are
relatively deep as evidenced by the larger height difference between
the $\tau_c = 1$ and $\tau_l = 1$ curves (given the assumption of LTE
intensity formation and an upward decreasing temperature).
In the intergranular lanes the temperature gradient is much shallower,
resulting in a lower sensitivity of the filter signal to temperature.
This is particularly clear at $x = 6$ arcsec, but also in the lanes
just outside strong magnetic field configurations at $x = 2.5$, and
4.4 arcsec.
%
%
%

From the position of the vertical dotted lines marking the location of
bright points in the filter intensity it is clear that these bright points
result from considerable weakening of the CH and CN lines.
At each of the bright point locations the $\tau_l = 1$ curve dips
down steeply along with the upward extension of the response function,
bringing the formation heights of the line cores closer
to those of the continuum, therefore weakening the line,
and amplifying the wavelength integrated filter signal.
This dip, which is the result of the partial evacuation of the
magnetic elements, is more pronounced in the CN line-opacity
because CN number densities decrease with depth in the flux concentration.
(see Figure \ref{fig:concentrations} and Section \ref{sec:densities}).

Remarkably, the CN band head proper with many overlapping lines has such
a high opacity that it forms considerably higher than
typical CN lines in the 388.3\,nm interval (see the dash-dotted curve
in the top panel in Figure \ref{fig:response}).
This means that its emergent intensity is less sensitive to the magnetic
field presence because the field in the higher layers is less concentrated
and, therefore, less evacuated, leading to a less pronounced dip
in the optical depth $\tau_h = 1$ curve.
Narrow band filtergrams or spectroheliograms
      \citep[e.g., ][]{Sheeley1971}
that mostly cover the CN 388.3\,nm band head can therefore be expected
to have less contrast than filtergrams obtained through the
1\,nm wide filters typically used in observations.

%
%

\section{RMS intensity variation and
bright-point contrast\label{sec:contrast}}
%
%
\begin{figure}[htbp]
  \epsscale{0.4}
  \plotone{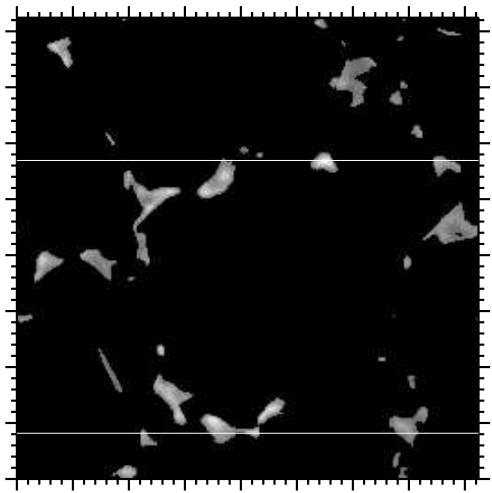}
  \caption{The G-band bright-point mask applied to the CH filtergram.
           The horizontal lines mark the same cross sections as in
           Figure \ref{fig:maps}.\label{fig:mask}}
  \epsscale{1.0}
\end{figure}
%
We now turn to a comparison of the relative filter intensities and
bright-point contrasts in the synthetic CH and CN band filtergrams.
To study the properties of bright points in the two
synthetic filtergrams we need to isolate them.
In observations this can be done by subtracting a continuum image,
in which the bright points have
lower contrast but the granulation has similar contrast,
from the molecular band image
      \citep{Berger+Loefdahl+Shine+Title1998a}.
We employ a similar technique here, but instead of using an
image obtained through a broad-band filter centered on a region
relatively devoid of lines, we use an image constructed from
just one wavelength position, namely at $\lambda = 430.405$\,nm.
More specifically, we count all pixels that satisfy
\begin{equation}
   \frac{f_{\mathrm{G-Band}}}{<f_{\mathrm{G-Band}}>} - 0.65 \frac{
     I_{430.405}}{<I_{430.405}>}\quad \geq 0.625
   \label{eq:mask}
\end{equation}
as bright point pixels, where $f_{\mathrm{G band}}$ is the G-band
filtergram, $I_{430.405}$ the continuum image, and averaging is
performed over the whole FOV.
The value 0.65 was chosen to optimally eliminate granular contrast
in the difference image, while the limit 0.625 was chosen so that
only pixels with a relative intensity larger than 1.0 were selected.
Furthermore, we define the average quiet-Sun intensity
$<f>_{QS}$ as the average of $f$ over all pixels that form
the complement mask of the bright-point mask.
The resulting bright-point mask is shown in Figure~\ref{fig:mask}
applied to the G-band filtergram.
\begin{table}[h]
  \caption{RMS intensity variation and average bright-point contrast
           in the synthetic CH and CN filtergrams.
           \label{tab:contrasts}}
  \vspace*{2ex}
  \begin{tabular}{lcc}
     \hline\hline
     Filter        &   RMS intensity   &  BP contrast \\ \hline
     G-band        &     20.5\%        &     0.497 \\
     CN            &     22.0\%        &     0.478 \\
     CN (Zakharov) &     19.7\%        &     0.413 \\
     CN (SST)      &     23.6\%        &     0.461 \\ \hline
  \end{tabular}
\end{table}
Defining the contrast of a pixel as
\begin{equation}
  C = f / <f>_{QS} -1,
\end{equation}
the bright point mask is used to compute the
contrast of bright points in the CH and CN filtergrams
for the different filters listed in Table \ref{tab:filters}.
The results are presented in Table \ref{tab:contrasts} along
with the RMS intensity variation over the whole FOV.
The synthetic filtergram CN filter centered at 388.3\,nm
yields an average bright-point contrast of 0.478,
very close to the value of 0.481 reported by 
      \citet[][their table 1]{Zakharov+Gandorfer+Solanki+Loefdahl2005}.
We find an average bright point contrast for the CH filter of
$<C_{\mathrm{CH}}> = 0.497$, which is much higher than the experimental
value of 0.340 reported by these authors.
Averaged over all the bright points defined in Eq. [\ref{eq:mask}]
we find a contrast ratio of $<C_{\mathrm{CN}}> / <C_{\mathrm{CH}}> = 0.96$,
in sharp contrast to the observed value of 1.4 quoted by
      \citet{Zakharov+Gandorfer+Solanki+Loefdahl2005}.
Using their filter parameters, moreover,
we find an even lower theoretical value of
$<C_{\mathrm{CN}}> = 0.413$, and a contrast ratio of only 0.83.

This variation of bright-point contrast in the CN filter filtergrams with
the central wavelength of the filter is caused by the difference in
the lines that are covered by the filter passband.
In the case of the La Palma filter and the Zakharov filter in particular,
the filter band integrates over several strong atomic lines,
which are less susceptible to line weakening than the molecular lines,
and therefore contribute less to the contrast enhancement of
magnetic elements (see Figure \ref{fig:detailspectra} in the next section).

%
%
\begin{figure}[htbp]
  \epsscale{0.65}
  \plotone{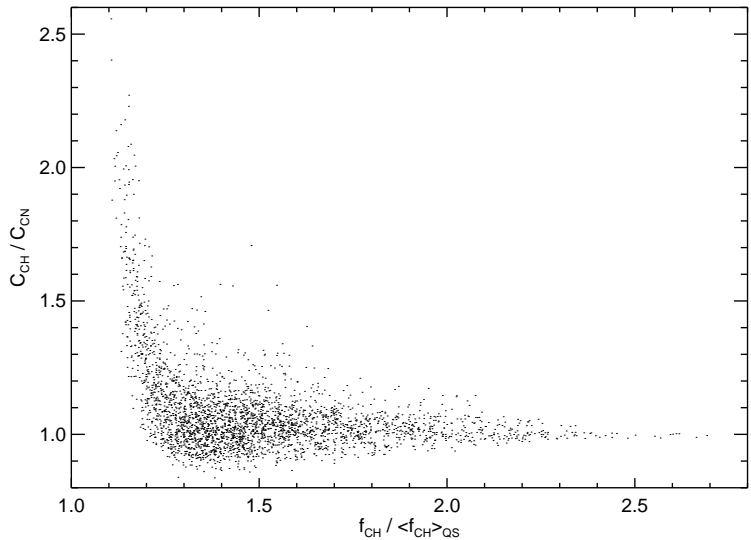}
  \caption{Scatter plot of the CH and CN band contrasts of bright-point
           pixels versus relative intensity in the CH filtergram.
           \label{fig:scatter}}
  \epsscale{1.0}
\end{figure}
Figure \ref{fig:scatter} shows the scatter in the ratio of
CH over CN contrast for all individual bright-point pixels.
At low CH intensity values of $f / <f>_{QS} < 1.3$ the CH contrast
is much larger than the contrast in the CN filtergram.
Above this value the contrast in CH and CN is on average very similar
with differences becoming smaller towards the brightest points.
Note that the scatter of the contrast ratio in CH and CN is not
dissimilar to the one presented by
      \citet[][their figure 4]{Zakharov+Gandorfer+Solanki+Loefdahl2005}
except that the label on their ordinate contradicts the conclusion
in their paper and appears to have the contrast ratio reversed.
%
%
\begin{figure}[bhp]
  \epsscale{0.7}
  \plotone{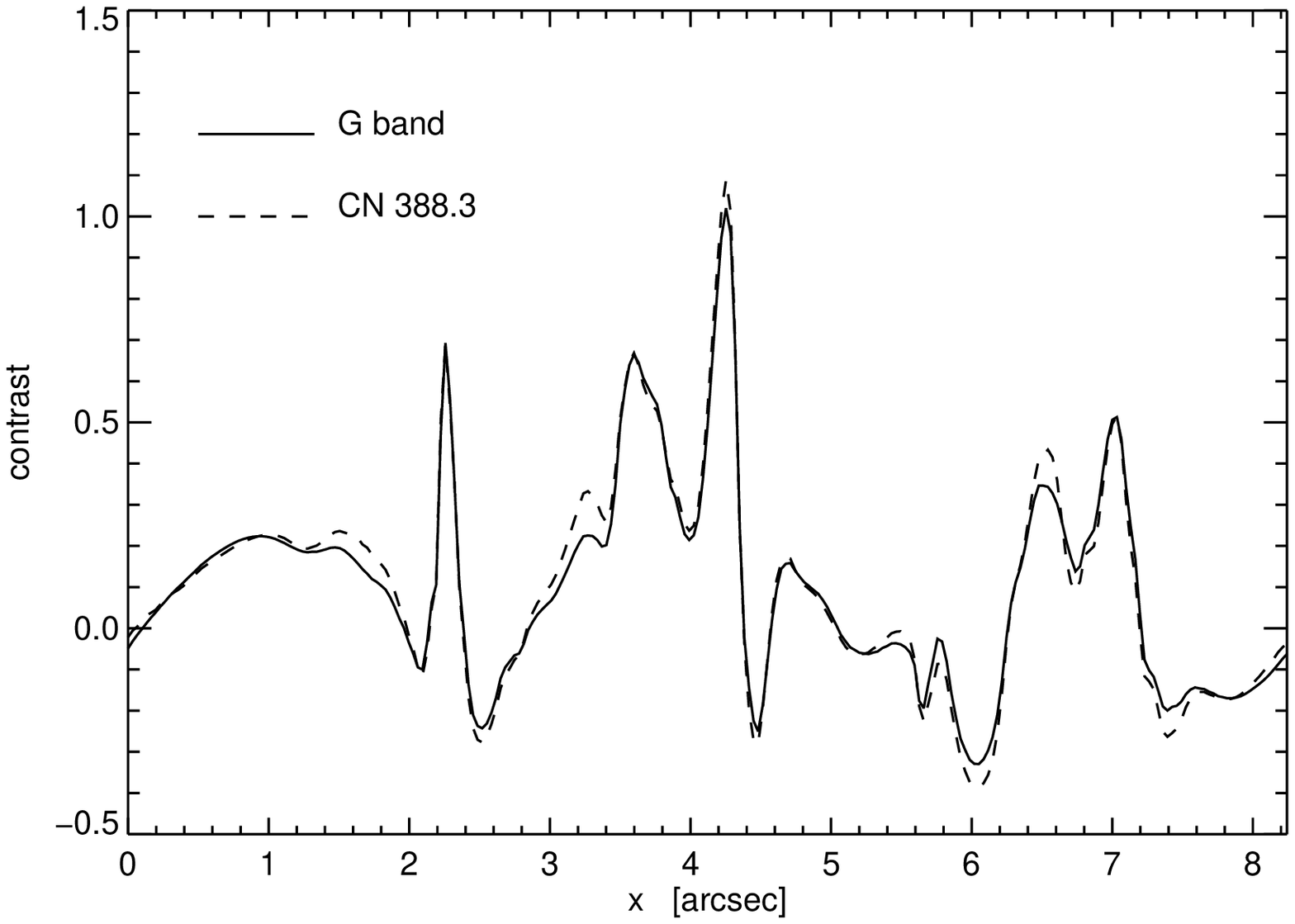}
  \plotone{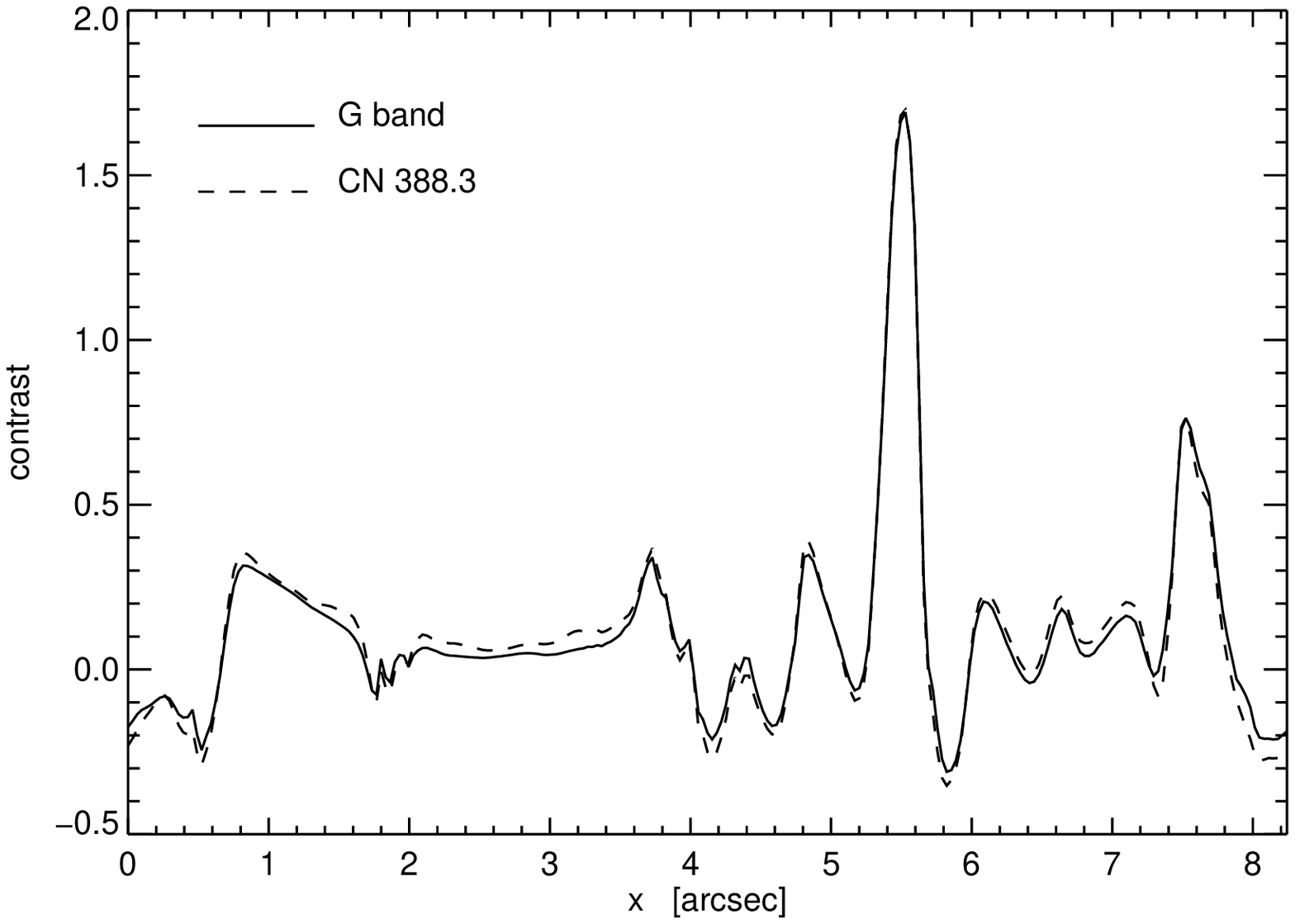}
  \caption{Relative intensities of the CH (solid curves) and
           CN (dashed curves) filter images in two cross sections
           of the simulation snapshot indicated by the horizontal
           lines in the left panel of Figure \ref{fig:maps}
	   (bottom panel corresponds to cross section $a$,
           the same cross section was used for the response function
           in Figure \ref{fig:response}),
           to panel corresponds to $b$.\label{fig:cuts}}
  \epsscale{1.0}
\end{figure}
%

To better display the difference in contrast between the two filtergrams
we plot their values in two cross sections indicated by the
horizontal lines in the left panel of Figure \ref{fig:maps} in
Figure \ref{fig:cuts}.
The contrast is clearly higher in granules and lower in intergranular
lanes in the CN image, but is identical in the bright points
(at $x = 2.2$, 3.6, 4.2, and 7.0\,arcsec in the left panel, and at
$x = 3.8$, 5.6, and 7.6 arcsec in the right panel, see also
Figure \ref{fig:mask}).
The lower contrast in the lanes and higher contrast in the granules
in CN is caused by the higher sensitivity of the Planck
function at the shorter wavelength of the CN band head when compared
to the G band. 

%
%

\section{Discussion}\label{sec:discussion}
In the synthetic CH- and CN-band filtergrams we find an average
bright-point contrast ratio $<C_{\mathrm{CN}}> / <C_{\mathrm{CH}}> = 0.96$
which is very different from the observational value of 1.4 reported by
      \citet{Zakharov+Gandorfer+Solanki+Loefdahl2005}.
If we employ the parameters of the CN filter specified by these
authors with a central wavelength of 388.7\,nm, redward of the
CN band head, we find an even lower theoretical contrast ratio 0.83.
Previously, several authors
      \citep{Rutten+Kiselman+Rouppe+Plez2001,%
Berdyugina+Solanki+Frutiger2003}
have predicted, on the basis of semi-empirical fluxtube modeling,
that bright points would have higher contrast in the CN-band with contrast
ratio values in line with the observational results of 
      \citet{Zakharov+Gandorfer+Solanki+Loefdahl2005}.
In these semi-empirical models it is assumed that flux elements
can be represented by either a radiative equilibrium atmosphere
of higher effective temperature, or a hot fluxtube atmosphere
with a semi-empirically determined temperature stratification,
in which case the stronger non-linear dependence of the Planck function
at short wavelengths results in higher contrast in the CN band.
Indeed if we use the same spectral synthesis data as for
the three-dimensional snapshot, and define the ratio of contrasts
in the CN band over the CH Band as
\begin{equation}
  R = \frac{f_{\mathrm{CN}}(T_{\mathrm{eff}})/f_{\mathrm{CN}}(5750) - 1.0}{
       f_{\mathrm{CH}}(T_{\mathrm{eff}})/f_{\mathrm{CH}}(5750) - 1.0},
\end{equation}
where $f(T_{\mathrm{eff}})$ is the filter signal for a Kurucz model
with effective temperature $T_{\mathrm{eff}}$,
we find that $R$ increases to 1.35 for $T_{\mathrm{eff}} = 6250$
and then decreases again slightly for higher effective temperatures because
the CN lines weaken more than the CH lines
      \citep[see also][]{Berdyugina+Solanki+Frutiger2003}.

However, more recent modeling, using magnetoconvection simulations like
the one employed here has shown that magnetic elements derive
their enhanced contrast from the partial evacuation in high field
concentrations, rather than from temperature enhancement
     \citep{Uitenbroek2003,Keller+Schuessler+Voegler+Zacharov2004,%
Carlsson+Stein+Nordlund+Scharmer2004}.
Here we make plausible that the close ratio of bright-point contrast
in CN and CH filtergrams we find in the synthetic images is
consistent with this mechanism of enhancement through evacuation.
%
%
\begin{figure}[tbph]
  \epsscale{0.75}
  \plotone{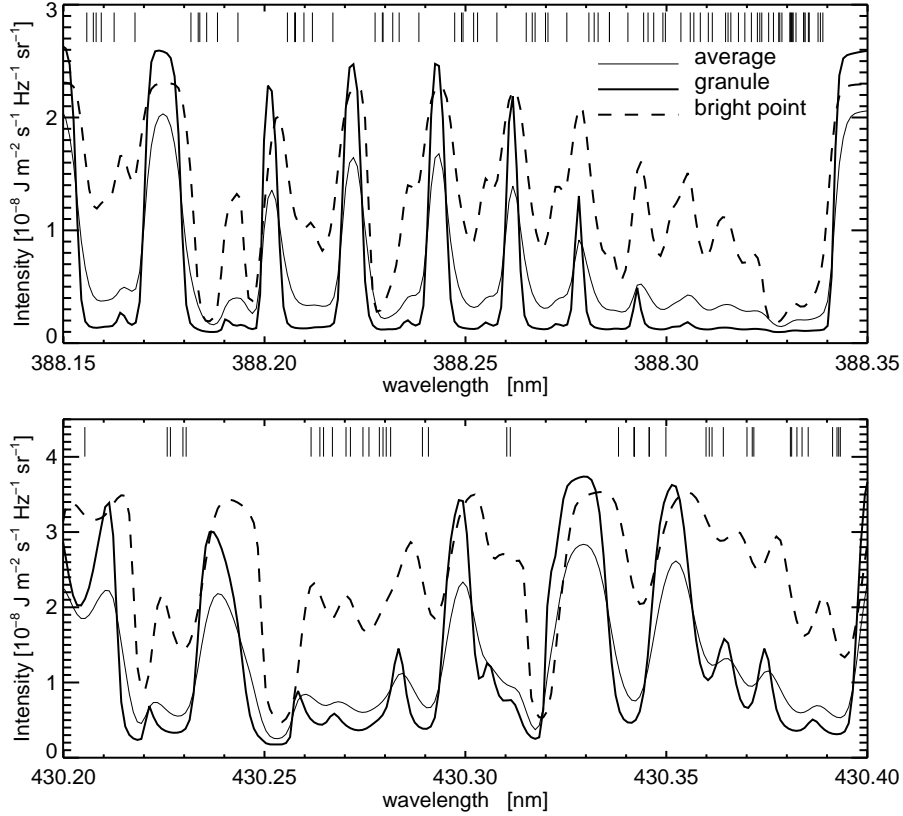}
  \caption{Short sections of G-band (bottom) and CN band (top) spectra
           for a typical granule (thick solid curve), bright point
           (thick dashed), and the spatial average (thin solid).
           Vertical lines at the top mark positions of CH and CN lines
           in the two intervals, respectively. 
           \label{fig:detailspectra}}
  \epsscale{1.0}
\end{figure}
%

Analysis of the filter response function to temperature,
and the behavior of the formation height of lines and the continuum in the
CN- and CH-band as traced by the curves of optical depth unity (see Figure
\ref{fig:response}) already indicate that the evacuation of magnetic
elements plays an important role in the appearance of these structures
in the filtergrams.
This is even more evident in the short sections of spectra
plotted in Figure \ref{fig:detailspectra},
which show the average emergent intensity over the
whole snapshot (thin solid curve), and the intensity from a bright
point (thick dashed curve) and a granule (thick solid curve) on an
absolute intensity scale.
Comparing the granular spectrum with that of the bright-point we
notice that their continuum values are almost equal but that the line
cores of molecular lines have greatly reduced central intensities
in the bright point, which explains why the magnetic structures can
become much brighter than granules in the CN and CH filtergrams.
If the high intensity of bright points in the filtergrams would arise
from a comparatively higher temperature, also their continuum intensities
would be higher than in granules.
Observational evidence for weakening of the line-core intensity
in G-band bright points without brightening of the continuum
is provided by
       \citet{Langhans+Schmidt+Rimmele+Sigwarth2001,%
Langhans+Schmidt+Tritschler2002}.
%
%
\begin{figure}[htbp]
  \epsscale{0.75}
  \plotone{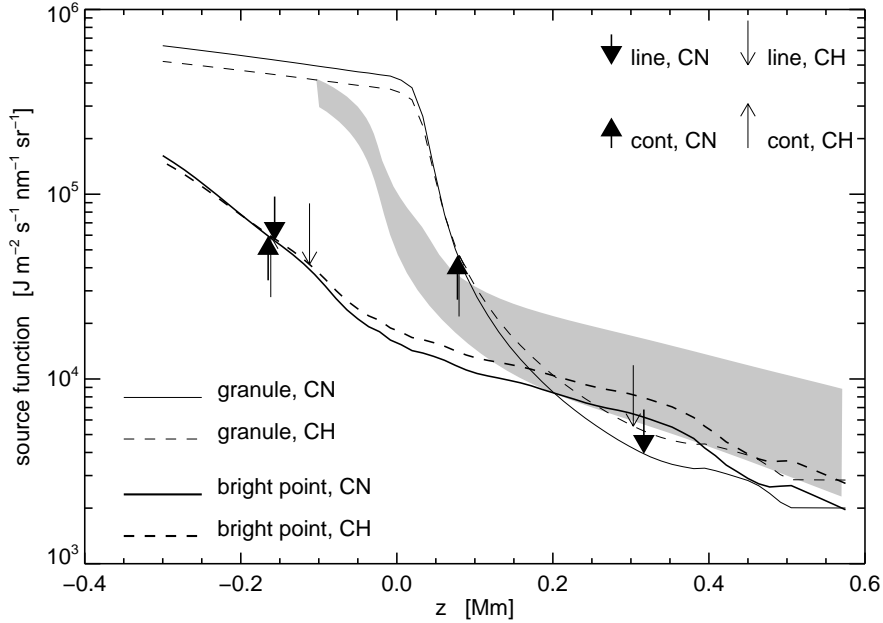}
  \caption{Source function of typical granule (thin curves) and
           bright point (thick curves) for the CH (dashed) and
           CN band (solid).
           Upward arrows mark the location of optical depth unity for
           continuum wavelengths,
           downward arrows that for line-center wavelengths.
	   The shaded area marks the region between the Planck functions
           for solar Kurucz models of effective temperature
           5750\,K and 6750\,K.\label{fig:temptau}}
  \epsscale{1.0}
\end{figure}
%

Line weakening in Figure \ref{fig:detailspectra} is less pronounced
in the CN band head at
388.339\,nm because the overlap of many lines raises the formation
height to higher layers where the density is less affected by evacuation
(see Sect.\ \ref{sec:response}).
Also atomic lines are less affected by the evacuation than lines of
the CN and CH molecule (e.g., compare the lines at $\lambda =
430.252$\,nm and 430.320\,nm with the weakened CH lines in the bottom
panel of Figure \ref{fig:detailspectra}), because the concentration
the atomic species is only linearly dependent on density, while that
of diatomic molecules is proportional to the square of the density of
their constituent atoms.
The latter effect is clear in the reduced number densities 
of CN and CH in bright points compared to intergranular lanes as
shown in Figure \ref{fig:concentrations} (see Section \ref{sec:densities}).

The partially evacuated magnetic concentrations
are cooler than their surroundings in a given geometric layer.
Radiation however, escapes both regions from similar temperatures,
at necessarily different depths.
This is made clear in Figure \ref{fig:temptau}, which shows
the source function (i.e., the Planck function, since we assume LTE)
in the CH and CN bands for the location of the same granule and
bright point for which the spectra in Figure \ref{fig:detailspectra}
are drawn.
The upward arrows mark the location of optical depth unity in the local
continua, and the downward arrows mark the same for typical
CN and CH lines in the bands.
Both continuum and the CN and CH line centers in the bright point
form approximately 250\,km below the continuum in the granule,
and they form very close together in both bands,
resulting in pronounced weakening of the molecular lines.
The structure of the response function (Figure \ref{fig:response})
indicates that the continuum contributes dominantly to the temperature
sensitivity of the filter integrated signals.
It forms almost at the same temperature in the bright point and granule, 
hence the comparable continuum intensities in Figure \ref{fig:detailspectra},
and the comparable brightness of granules and magnetic elements in
continuum images.
It is precisely for this reason that the bright-point contrast in
the synthetic CN filtergram is very similar to that in CH,
instead of being much higher.
The high contrast of magnetic concentrations in these filtergrams
results from line weakening in the filter passband,
and not from temperatures that are higher than in a typical granule
at the respective formation heights of the filter integrated signal.

The shaded region in Figure \ref{fig:temptau} indicates the range
between Planck functions for solar Kurucz models
     \citep{Kurucz_CD13}
of effective temperatures between 5750\,K (bottom) and 6750\,K
(top) at the central wavelength of the G band filter.
The comparison with the granule and bright point source functions
shows that neither can be represented by a radiative equilibrium model,
except near the top, where the mechanical flux in the simulations vanishes,
and where the temperature in both structures converges towards the
temperature in the standard solar model of $T_{\mathrm{eff}} = 5750$\,K.
In particular, the temperature gradient in the flux element
is much more shallow as the result of a horizontal influx of
radiation from the hotter (at equal geometric height) surroundings.
This shallow gradient further contributes to the weakening of
the molecular spectral lines.

%
%

\section{Conclusions}\label{sec:conclusion}
We have compared the brightness contrast of magnetic flux concentrations
between synthesised filtergrams in the G-band and in the violet CN band at 
388.3\,nm, and find that, averaged over all bright points in the
magnetoconvection simulation, the contrast in the CN band is lower
by a factor of 0.96.
This is in strong contradiction to the observational result reported by
      \citet{Zakharov+Gandorfer+Solanki+Loefdahl2005},
who find that the bright-point contrast is typically 1.4 times higher
in CN-band filtergrams.
In the present simulation the enhancement of
intensity in magnetic elements over that of quiet-Sun features
is caused by molecular spectral line weakening in the partially
evacuated flux concentration.
At the median formation height of the filter intensity 
(as derived from the filter's temperature response function,
Figure \ref{fig:response}) the temperature in the flux concentration
is comparable to that of a typical granule (Figure \ref{fig:temptau}).
As a result of these two conditions the contrast between the
bright point intensity and that of the average quiet-Sun is very
similar in both the CH and CN filters, and not higher in the
latter as would be expected from Planck function considerations
if the enhanced bright point intensity were the result of
a higher temperature in the flux concentration at the filter
intensity formation height.

The ratio of CH bright point contrast over that of the CN band 
varies with bright point intensity (Figure \ref{fig:scatter}),
with a relatively higher G-band contrast for fainter elements.
Theoretically, this makes the G band slightly more suitable
for observing these lower intensity bright points.

Because the bright-point contrast in filtergrams is the result of
weakening of the molecular lines in the filter passband its value
depends on the exact position and width of the filter (see Table
\ref{tab:contrasts}).
The transmission band of the filter used by
     \citet{Zakharov+Gandorfer+Solanki+Loefdahl2005}
mostly covers atomic lines of neutral iron and the hydrogen Balmer
H$_8$ line, and hardly any CN lines because it was centered redward
of the band head at 388.339\,nm.
Hence the theoretical contrast obtained with this filter is even lower
than with the nominal filter centered at the band head.

We find that the RMS intensity variation in the CN filtergram is
slightly higher than in the CH dominated G band with values of 22.0\,\%
and 20.5\,\%, respectively.
The former value depends rather strongly on the central wavelength
of the employed filter (Table \ref{tab:contrasts}).
The greater intensity variation in the CN-band filtergram is
the result of the stronger temperature sensitivity of the Planck
function at 388.3\,nm than at 430.5\,nm.
These intensity variations are moderated by the fact that the
filter integrates over line and continuum wavelengths combined
with a decrease in horizontal temperature variation with height,
the strong opacity dependence of the H$^-$ opacity,
and the strong decrease of the CN number density with temperature and
depth in the intergranular lanes (Section \ref{sec:densities}).
The low RMS intensity variation through the filter described by
     \citet{Zakharov+Gandorfer+Solanki+Loefdahl2005}
is the result of the inclusion of the hydrogen H$_8$ line in the passband.
Similarly to the H$\alpha$ line the reduced contrast in the
H$_8$ line is the result of the large excitation energy of its lower
level, which makes the line opacity very sensitive to temperature
     \citep{Leenaarts_etal2005}.
A positive temperature perturbation will strongly increase the
hydrogen $n_2$ number density (through additional excitation in
Ly$\alpha$) forcing the line to form higher at lower temperature
thereby reducing the intensity variation in the line, and vice versa
for a negative perturbation.

Finally, the mean spectrum, averaged over the area of the simulation
snapshot, closely matches the observed mean disk-center intensity
(Figure \ref{fig:spectrum}), providing confidence in the realism of
the underlying magnetoconvection simulation and our numerical
radiative transfer modeling.
Moreover, the filter integrated quantities we compare here
are not very sensitive to the detailed shapes of individual
spectral lines (for instance, the filter contrasts are the same
with a carbon abundance $\epsilon_{C} = 8.60$, although the
CN lines in particular provide a much less accurate fit to the
mean observed spectrum in that case).
The clear discrepancy between the observed and synthetic contrasts,
therefore, indicates that we lack complete understanding of either the
modeling, or the observations, including the intricacies of image
reconstruction at two different wavelengths at high resolution,
or both.
In particular, the near equality of bright point contrast in the
two CN and CH bands is a definite signature of brightening through
evacuation and the concomitant line weakening.
If observational evidence points to a clear wavelength dependence
of the bright point contrast, it may indicate that the simulations
lack an adequate heating mechanism in their magnetic elements.

%

\acknowledgements
We are grateful to Bob Stein for providing the
three-dimensional magneto-convection snapshot.
This research has made use of NASA's Astrophysics Data System (ADS).


%
%


\end{document}